# Aerocapture: A Historical Review and Bibliometric Data Analysis from 1980 to 2023

Athul Pradeepkumar Girija [1][**][**]

[1]*School of Aeronautics and Astronautics, Purdue University, West Lafayette, IN 47907, USA*

**ABSTRACT**

Aerocapture is a technique which uses atmospheric drag to decelerate a spacecraft and achieve nearly fuel-free orbit insertion from an interplanetary trajectory. The present study performs a historical review of the field, and a bibliometric data analysis of the literature from 1980 to 2023. The data offers insights into the evolution of the field, current state of research, and pathways for its continued development. The data reveal a pattern in the rise of publications, followed by a period of stagnation, which repeats itself approximately once every decade. Mars is the most studied destination, while Uranus is the least studied. Prior to 2013, NASA centers produced the most publications and are the most cited in the field. However, academic institutions produced the majority of publications in the last decade. The United States continues to be the leading country in terms of publications, followed by China. The Journal of Spacecraft and Rockets is the leading source of publications, both in terms of number and citations. NASA is the leading funding source, followed by the National Natural Science Foundation of China. A proposed low-cost Earth flight demonstration of aerocapture will greatly reduce the risk for future science missions.

---

[****] To whom correspondence should be addressed, E-mail: athulpg007@gmail.com

# I. INTRODUCTION

Aerocapture is a technique which uses atmospheric drag to decelerate a spacecraft and achieve nearly fuel-free orbit insertion from an interplanetary trajectory. Compared to propulsive insertion which requires substantial amount of fuel, aerocapture can offer significant mass savings which can translate into more useful scientific payload or lower launch mass and cost. Figure 1 shows a schematic of the aerocapture mission concept. The vehicle flies through the atmosphere, protected by a heat-shield, exits the atmosphere, and then enters orbit around the planet. Aerocapture has never been flown on a planetary mission, but has been studied extensively over the last four decades for applications to every atmosphere-bearing destination in the Solar System. The present study performs a historical review of the field, and a bibliometric data analysis of the literature from 1980 to 2023 which offers interesting insights into the evolution of the field over the years, current state of research, and pathways for its continued development.

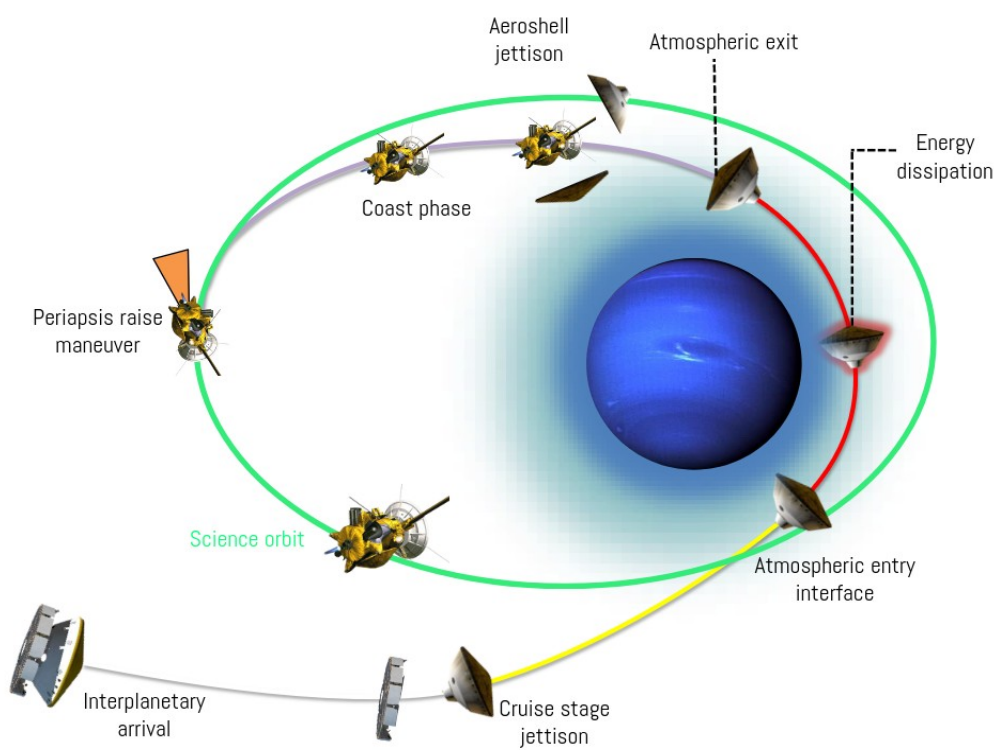

Figure 1. Concept of operations for the aerocapture mission concept.

## II. HISTORICAL REVIEW

The concept of using a planetary atmosphere to perform 'aerodynamic braking' for orbit insertion has been studied as early as 1965 by Finch and other authors [1, 2]. These studies then referred to the technique as 'aerobraking', which today refers to a different technique of using multiple upper atmospheric passes to slowly circularize the orbit of a spacecraft already in a highly elliptical orbit. The first use of the word 'aerocapture' is by Cruz in 1979, which is considered the seminal paper to articulate this concept [3]. The study initiated a series of studies in the early 1980s to further refine the vehicle and mission design for aerocapture concepts. In 1985, Walberg performed a comprehensive review of aero-assisted orbit transfer techniques including aerocapture [4]. Walberg predicted that 'practical aerocapture vehicles would be operational with a decade'. In the mid-to-late 1980s, the Aeroassist Flight Experiment (AFE) proposed a Shuttle-launched vehicle to demonstrate the end-to-end use of technologies used in a future aerocapture mission [5]. The AFE project led to several advances in guidance schemes, hardware and software development for the flight mission [6]. However, in the early 1990s the AFE project was eventually canceled to due to cost overruns thus precluding a flight demonstration of aerocapture.  Meanwhile, aerocapture mission concepts continued to be proposed in the early 1990s, most notably with applications to Mars Sample Return missions [7]. In the mid-1990s, aerocapture was initially baselined for the Mars 2001 orbiter mission, but was later dropped in favor or aerobraking due to budgetary constraints, and to reduce the risk following the failure of the Mars Polar Lander and the Mars Climate Orbiter missions [8]. In the late 1990s, aerocapture was considered for the joint NASA-CNES Mars Sample Return mission to be launched in 2007, and once again led to extensive research in guidance schemes and vehicle design, but the joint mission was eventually not flown [9].  In 2002, the Aerocapture Flight Test Experiment (AFTE) proposed a low-cost demonstration flight at Earth, one of the candidates under New Millennium Program (NMP) ST-7 competition but was not selected [10].  In the early 2000s, aerocapture was identified as one of the highest priority investments for the NASA In Space Propulsion Technology (ISPT) program [11]. The Aerocapture Systems Analysis Team (ASAT) formed as a part of this effort performed a detailed study of aerocapture concept and vehicle design for Venus, Mars, Titan, and Neptune missions [12, 13]. The ASAT studies represented a significant leap in our understanding of aerocapture system design and significantly influenced all future research in the field. In 2005, Hall et al. performed a cost-benefit analysis of the aerocapture mission set across the entire range of Solar System destinations and concluded aerocapture provides enormous advantages over propulsive insertion, especially for missions to the outer planets [14]. In 2006, an aerocapture technology demonstrator at Earth was once again proposed under the NMP ST-9 opportunity but was not selected [15].

Following the end of the ISPT program supporting aerocapture technology development in the mid-to-late 2000s, there was a lack of funding and significant research efforts in the field. Some mission concepts continued to be proposed such as the Titan Explorer which significantly leveraged the results of the ISPT studies [16]. While the ISPT studies showed that aerocapture provides enormous advantages for outer planet missions, the high risk associated with aerocapture for outer planet Flagship missions precluded it from being considered. In 2011, the 2013-2022 Planetary Science Decadal Survey stated "further risk reduction will be required before high value and highly visible missions will be allowed to utilize aerocapture techniques." Also, it was becoming increasingly clear that the high costs associated with lifting body aeroshells would preclude aerocapture missions in a cost-competitive environment. In 2013, Putnam and Braun published the concept of drag modulation aerocapture as a simpler flight control method which would obviate the need for complex control systems and potentially enable cheaper aerocapture missions [17]. Though drag modulation control techniques had been published prior to 2013, their study was the first to perform a comprehensive analysis of the technique and its applicability to low-cost missions. The study inspired several small and low-cost missions using aerocapture at Mars and Venus in the mid-to-late 2010s [18, 19]. In 2016, there was renewed interest in aerocapture in Uranus and Neptune in preparation for the Ice Giants Pre-Decadal Survey. A JPL study investigated the readiness of aerocapture and concluded that while no technology developments are needed for Mars, Venus, and Titan missions; significant technology developments would be required for a mid-L/D vehicle to accommodate the large uncertainties at the ice giants [20]. Further work showed that with improvements in approach navigation and atmospheric guidance schemes, it is possible to perform aerocapture at the ice giants using heritage blunt-body aeroshells and thus avoid the need for mid-L/D vehicles [21]. Meanwhile, in the late 2010s, a significant number of studies continued to be performed to assess the viability of inserting small satellites into orbit around Mars and Venus using drag modulation aerocapture. In 2019, Werner and Braun proposed a low-cost flight demonstration at Earth using a drag modulation vehicle [22]. In 2022, Girija et al. performed a quantitative assessment of the feasibility of lift and drag modulation control, and mass-benefit analysis of aerocapture across the Solar System [23, 24, 25]. The 2023-2032 Planetary Science Decadal concluded that "aerocapture is a technology that is ready for infusion and that can enhance or enable a large set of missions, but that will require special incentives by NASA to be proposed and used in a science mission" [26]. In 2023, Girija et al. showed how drag modulation aerocapture can enable large constellations of small satellites around Mars and Venus which fit within the envelope of low-cost missions and enable a new paradigm in planetary exploration [27]. Efforts continue to be made to realize a small low-cost drag modulation aerocapture technology demonstration mission at Earth to make it a more viable option for future missions [28].

## III. BIBLIOMETRIC ANALYSIS

The study performs a bibliometric analysis of the published literature on aerocapture. Bibiliometric studies can provide useful insights into how the field has evolved over time, its main research directions during various time periods, which institutions produce the most research output, and how changing funding priorities affect the field. The dataset was obtained by querying for the keyword 'aerocapture' from the Web of Science (WoS) Core Collection which returned about 270 relevant results. A few publications prior to 1980 were missing from theresults, and were manually added to the dataset. The total number of publications in the dataset is n=277, spanning the years from 1962 to 2023. The fields in the dataset include the following: Authors, Author full names, Authors' WoS unique identifier, Title, Year, Source title, Volume, Issue, Article No., Page Count, Citation count, Digital Object Identifier, Scopus link to the article, Author affiliations, Funding text, Publisher, Language, Document Type, and Data Source. A few derived fields were manually added to the dataset to aid in the analysis: Planet (derived from the title), Funding source (derived from funding text), Country of Origin (derived from affiliation), and Institution (derived from affiliation).

Figure 2 shows the cumulative publication count as a function of the year, which shows that the field has consistently grown over the past four decades and continues to grow. It is seen the concept existed in the literature as early as 1965, but Cruz's paper in 1979 kick started the growth. In highly established fields with substantial funding and large number of researchers, the number of publications typically grow exponentially, but in a relatively narrow field such as aerocapture the growth is well approximated by a quadratic curve fit (where x = years since 1980).

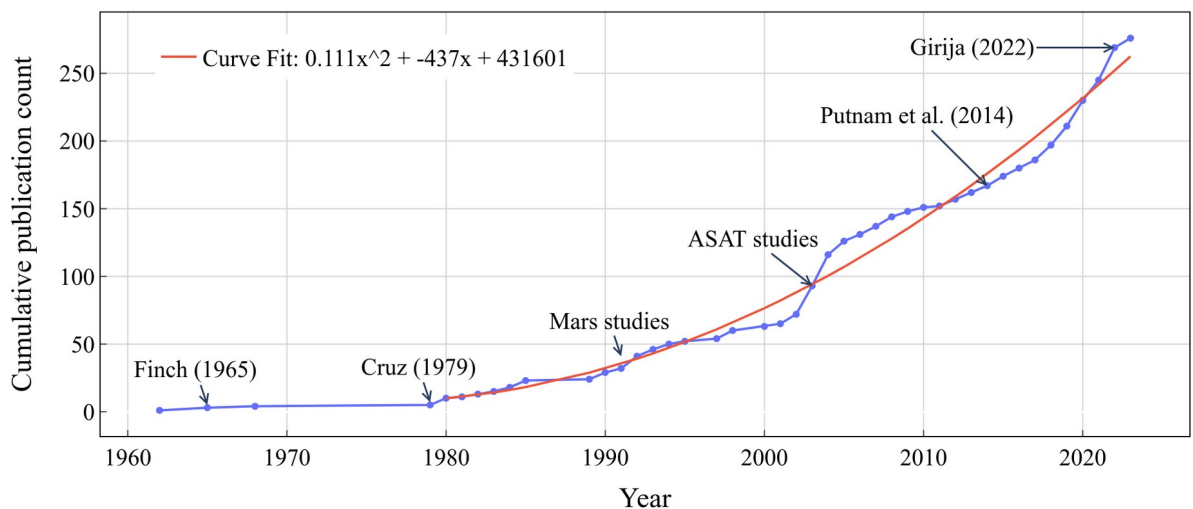

Figure 2. Cumulative count of aerocapture publications.

Figure 2 also shows that there is a pattern of a rising publications for a few years, followed by a period of stagnation, a cycle which repeats itself approximately once every 10 years. This is highlighted by how the data points are above the curve fit for a period of time, then fall below the curve, and then rise again. Of course, fitting a curve to any time series dataset with random variations would produce such an effect, but it is clear even without the curve fit that there are repeating periods of growth and stagnation. The curve fit only makes it easier to observe the repeating cycle. The growth phases are attributed to a periods where there is more research interest in aerocapture, supported by government funding which leads to more research output. The stagnation phases are when the funding from the growth phase runs out and researchers change their focus areas to those with more funding. A classic example of this is the ASAT studies in 2003, during which time NASA prioritized aerocapture technology and allocated substantial research investments which led to a rapid growth phase. By the late-2000s, however, aerocapture was no longer a top priority and hence funding became more difficult for researchers in the field, which led to a stagnation phase.

Figure 3 shows the number of publications per year, colored by target planet. Prior to 2000, most studies focused on Mars, with some generic design studies, and some Titan studies. Figure 3 also highlights the pattern of a few years with rising publications followed by a stagnation period which repeats itself approximately once every decade. The ASAT studies in 2003 produced one of the largest growth phases, followed by a decline in the late 2000s. Since 2016, a renewed interest in ice giant aerocapture, and small satellite aerocapture at Mars and Venus is driving the research output [29, 30]. The year 2022 produced the most publications, indicating the field may currently be at another peak.

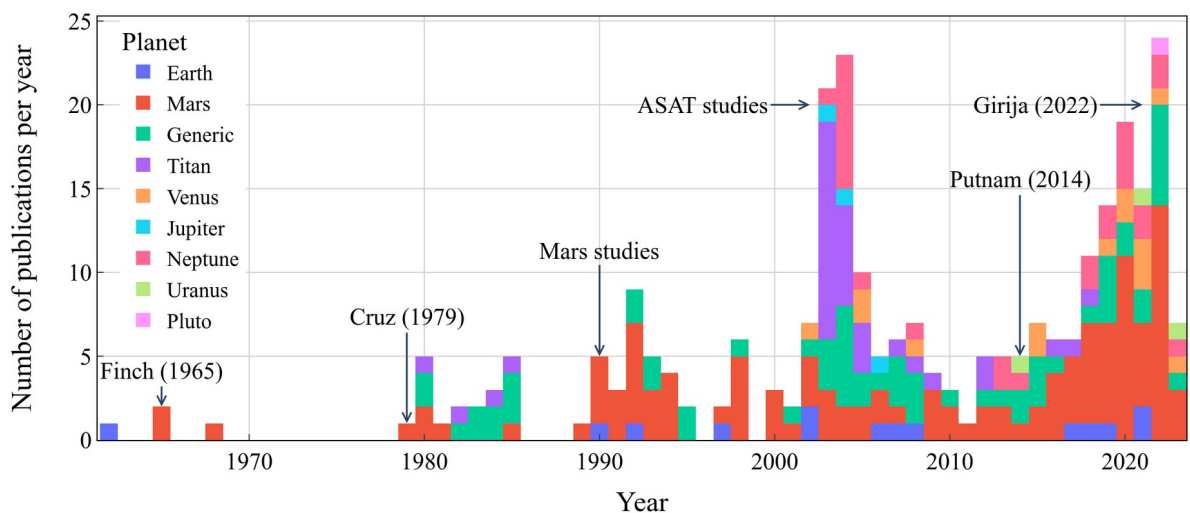

Figure 3. Number of publications per year colored by target planet.

Figure 4 shows the distribution of the target planets in the dataset, with the generic studies being excluded. With over half of all studies, Mars is the most studied destination for aerocapture, followed by Titan and Neptune. Uranus, which accounts for only 1% of the studies is the least studied planet. This can be attributed to the fact that during the ASAT studies, GRAM atmosphere models were developed for Mars, Titan, and Neptune but not for Uranus. UranusGRAM was released only in late 2020s which precluded detailed Uranus aerocapture studies until then [31].

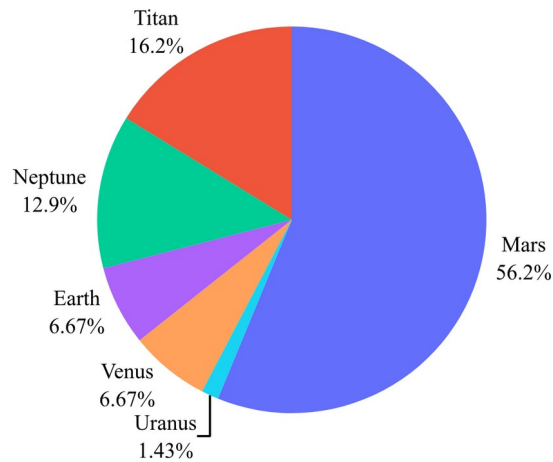

Figure 4. Distribution of the target planets in the dataset.

Figure 5 shows the distribution of institutions in the dataset over all time. The top institutions (in terms of number of publications) are NASA Jet Propulsion Laboratory (JPL), NASA Langley Research Center (LaRC), Purdue University, and University of Colorado, Boulder. This is due to the fact that much of the initial work in the 1980-2010 time period was done at NASA centers as seen in Figure 6, and only some publications from academic institutions.

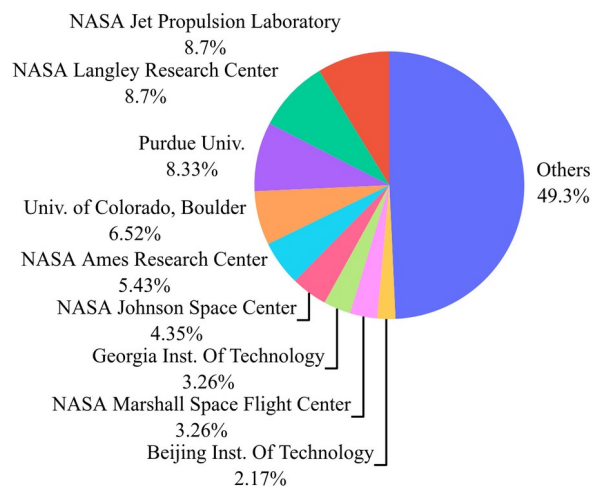

Figure 5. Distribution of the institutions in the dataset (all time).

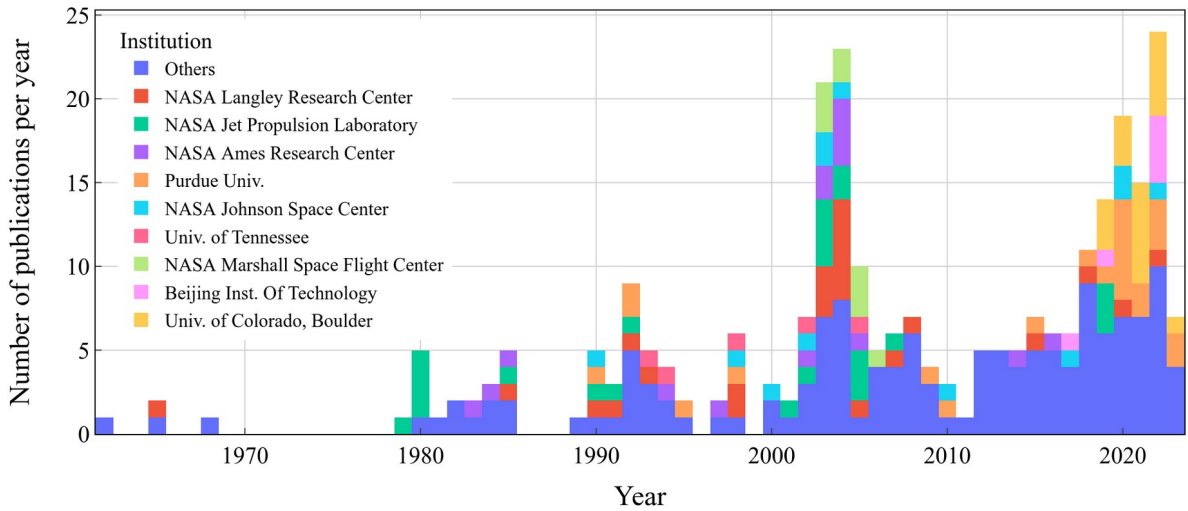

Figure 6. Number of publications per year colored by institution.

Figure 7 shows the distribution of institutions in the dataset in the last 10 years (since 2013). Compared to Figure 5, it is seen that academic institutions such as UC Boulder, and Purdue Univ. produced the majority of aerocapture publications in the last 10 years. This is attributed to change in the funding scenario. In the 1980s through the early 2000s, there was considerably more support for aerocapture research at NASA centers. However, in the last 10 years, there is a comparative lack of funding opportunities for NASA centers to work on aerocapture as it is not considered a high priority. Much of the available funding is through NASA fellowships and grants made to academic institutions.

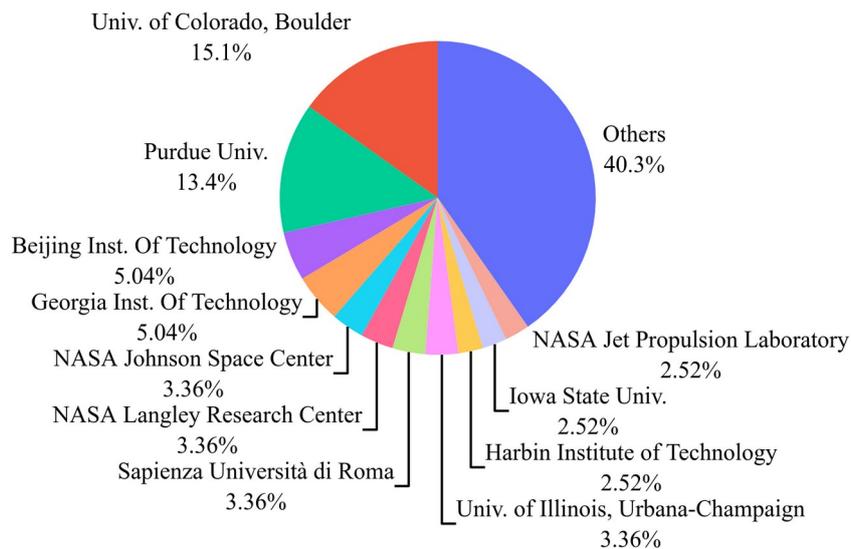

Figure 7. Distribution of the institutions in the dataset (last 10 years).

Figure 8 shows the distribution of country of origin in the dataset over all time. The United States is the leading country with 78% of all publications, followed by China, France, Italy, and Japan. Much of the initial work prior to 2000 was done almost exclusively in the United States as seen in Figure 9. Only after 2000, there is some research from other countries such as France and Japan. Figure 10 which shows the distribution of the country of origin in the last 10 years. The United States continues to be the leading source of publications, followed by China. It is noteworthy that China's share of publications has grown from essentially zero to 13% only in the span of a decade, which will be shown later to be largely driven by funding from the National Natural Science Foundation of China.

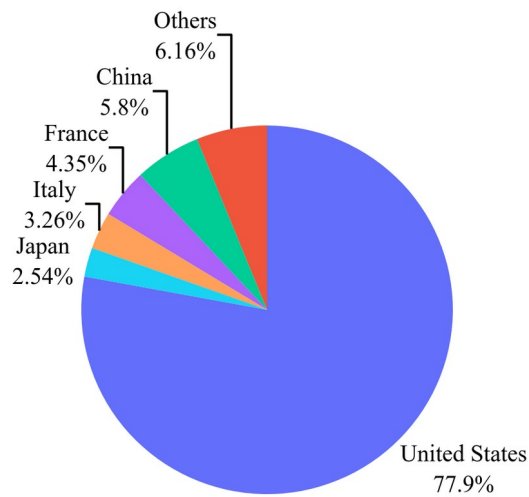

Figure 8. Distribution of country of origin in the dataset (all time).

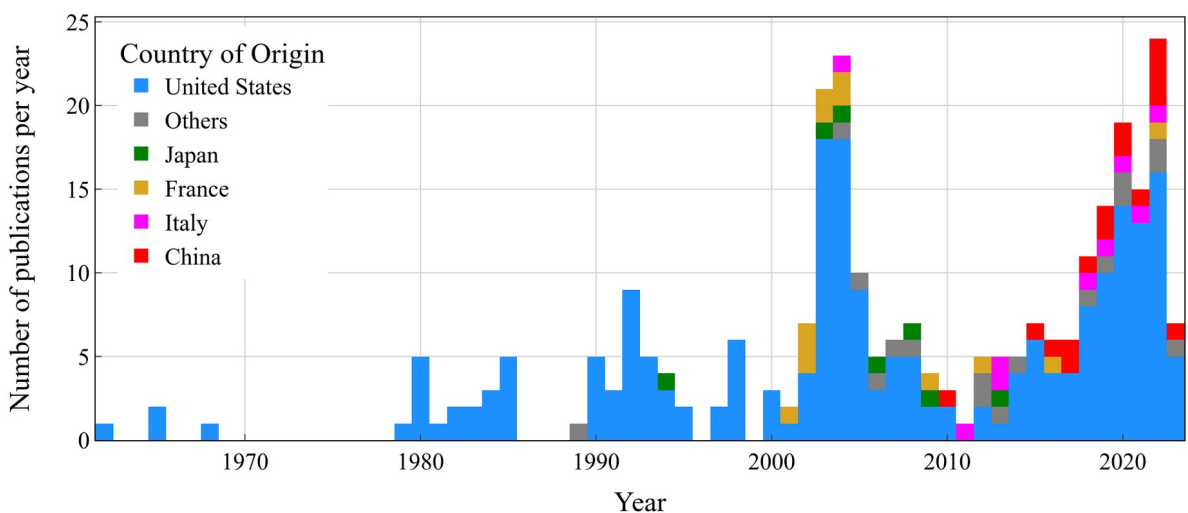

Figure 9. Number of publications per year colored by country of origin.

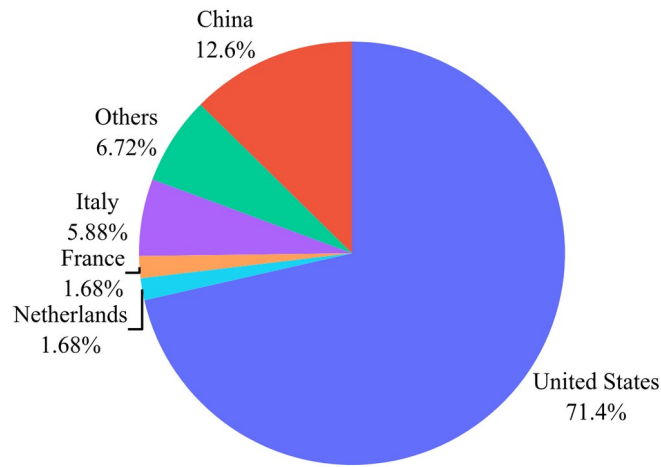

Figure 10. Distribution of country of origin in the dataset (last 10 years).

Figure 11 shows the distribution of source titles in the dataset. The Journal of Spacecraft and Rockets is the leading source of aerocapture publications, followed by Advances in the Astronautical Sciences, IEEE Conference Proceedings, and AIAA Scitech Conference Proceedings. AIAA Journal of Spacecraft and Rockets has historically been the most leading source for publications as evidenced by several key studies such as Hall et al. [14], Putnam and Braun [17], Spilker et al. [20], and Girija et al. [25] being published in the journal over the past two decades.

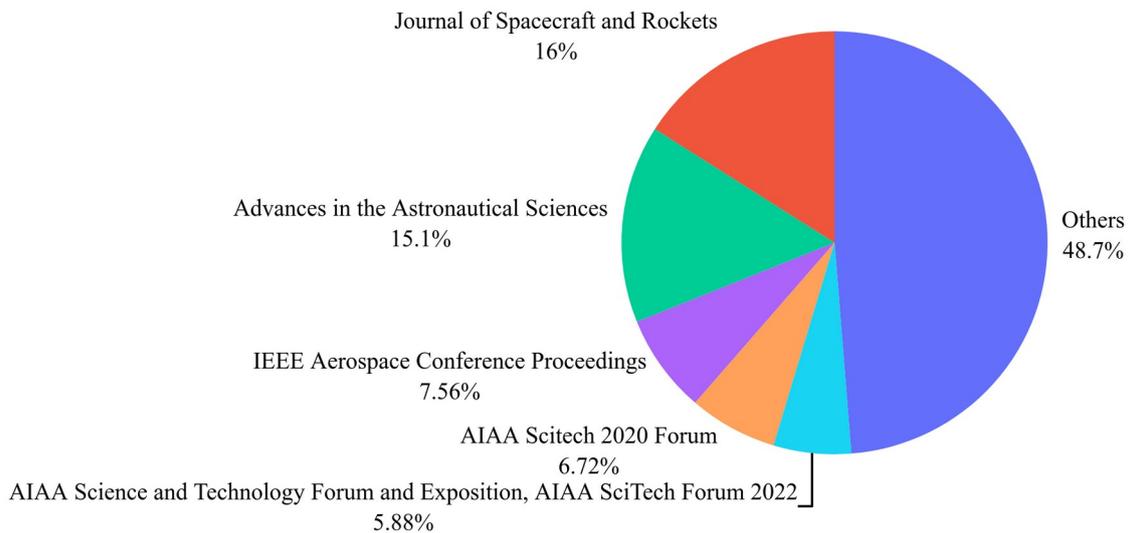

Figure 11. Distribution of source titles in the dataset.

Figure 12 shows the authors and citation counts of the ten most cited aerocapture studies of all time. The most cited study is the aero-assist review article by Walberg in 1985, followed by London's satellite plane change paper from 1962. Review articles and seminal papers such as these are often the most cited articles in any field. The third most cited is Hall et al. (2005) which was the first study to perform a comprehensive cost-benefit analysis of aerocapture at all destinations in the Solar System, and the mass benefit calculation from the study is widely referenced. Table 1 shows the year of publication, authors, title, and citation counts of the top ten most cited articles.

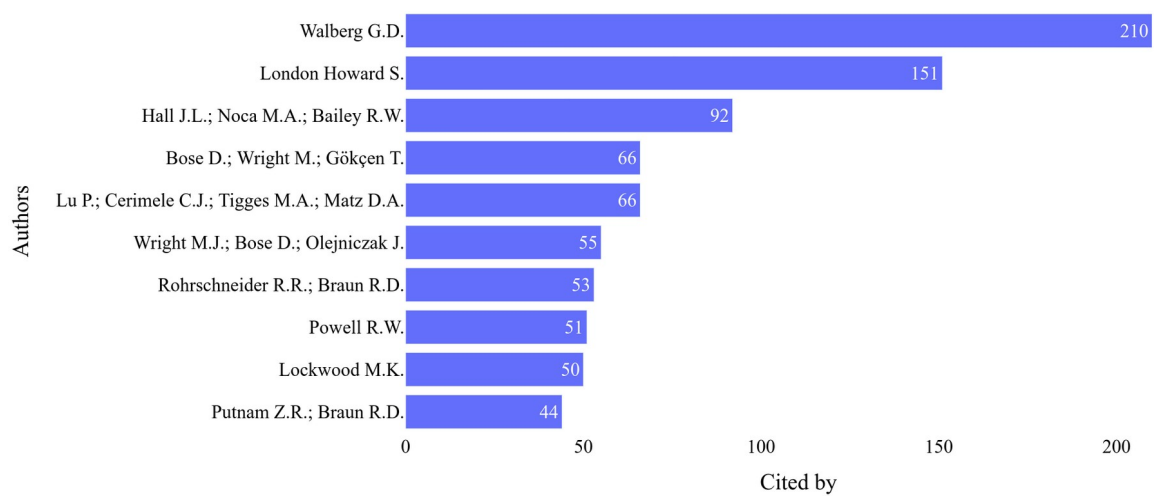

Figure 12. Citation counts of top 10 most cited studies of all time.

Table 1. Top ten most cited articles

| Year | Authors | Title | Count |
|------|---------|-------|-------|
| 1985 | Walberg | A survey of aeroassisted orbit transfer | 210 |
| 1962 | London | Change of Satellite Orbit Plane by Aerodynamic Maneuvering | 151 |
| 2005 | Hall et al. | Cost-benefit analysis of the aerocapture mission set | 92 |
| 2004 | Bose et al. | Thermochemical modeling for titan atmospheric entry | 66 |
| 2015 | Lu et al. | Optimal aerocapture guidance | 66 |
| 2005 | Wright et al. | Impact of flowfield-radiation coupling on aeroheating for Titan | 55 |
| 2007 | Rohrschneider et al. | Survey of ballute technology for aerocapture | 53 |
| 1998 | Powell | Numerical roll reversal predictor corrector aerocapture guidance | 51 |
| 2003 | Lockwood | Titan aerocapture systems analysis | 50 |
| 2014 | Putnam and Braun | Drag-modulation flight-control system options for aerocapture | 44 |

Figure 13 shows the top ten citation counts of the articles grouped by the source title. Journal of Spacecraft and Rockets (JSR) articles are the most cited, with a clear lead over the next most cited source which are AIAA conference papers. This shows that JSR articles have had the most impact in the field, measured in terms of citation count.

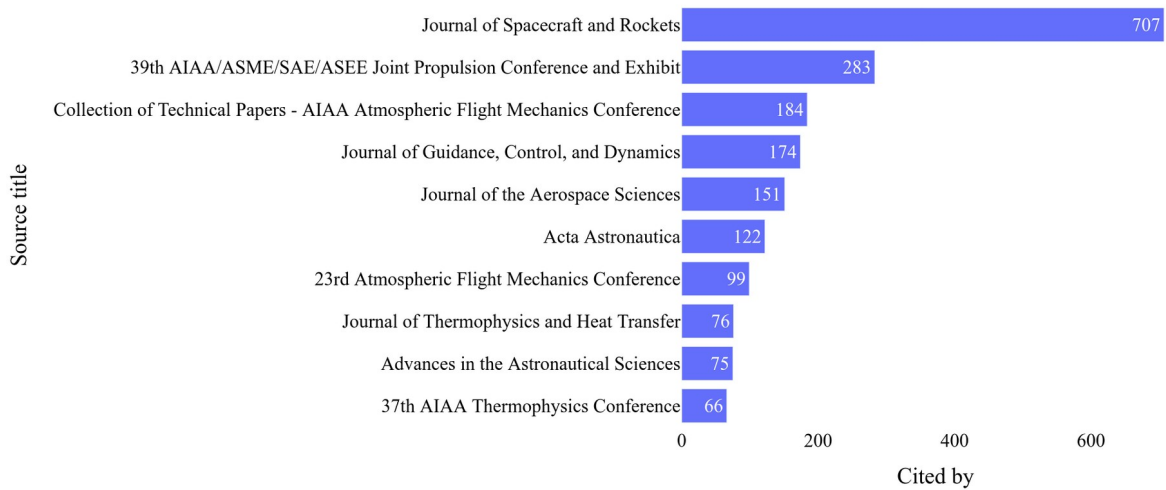

Figure 13. Citation counts grouped by source title.

Figure 14 shows the top ten citation counts of all time grouped by the institution. NASA LaRC, JPL, and Ames are the most cited institutions, followed by academic institutions such as Georgia Tech and Purdue University. This is again due to the fact that much of the early research in aerocapture until the early 2000s was done at NASA centers.

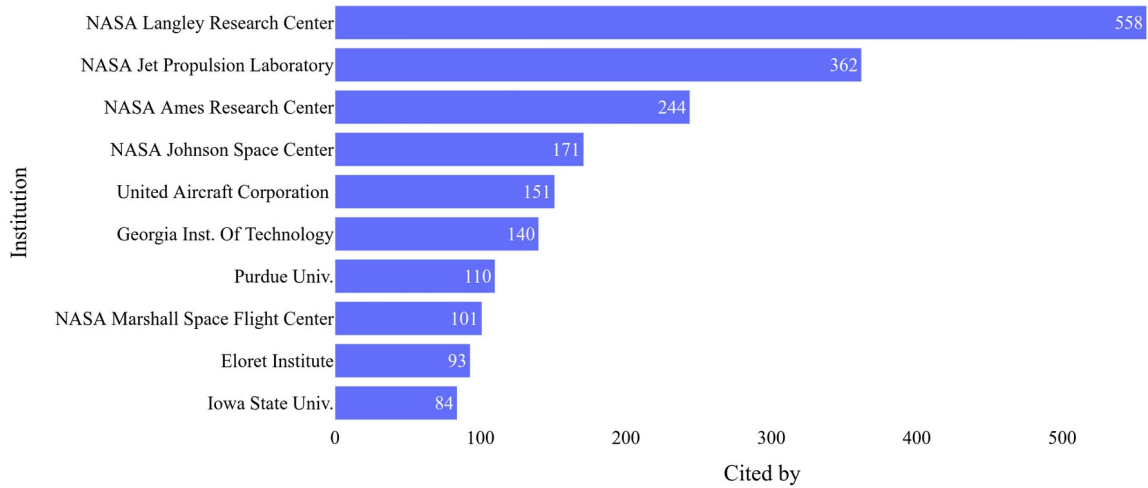

Figure 14. Citation counts of articles (all time) grouped by institution.

Figure 15 shows the top ten citation counts of articles in the last 10 years grouped by the institution. As with the number of publications shown in Fig. 7, the most cited institutions are academic in the last 10 years. The most cited institutions are Iowa State Univ., Georgia Tech. and Purdue Univ, while NASA centers have become less cited.

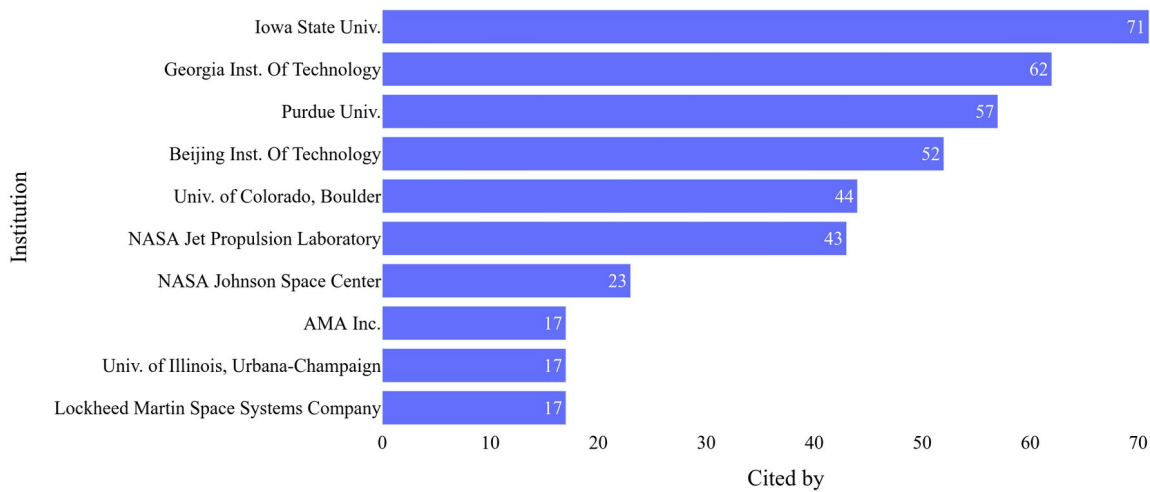

Figure 15. Citation counts of articles in the last 10 years grouped by institution.

Figure 16 shows the top three funding sources (for articles with funding text) prior to 2013. NASA is the leading funding source for aerocapture research, followed by much smaller contributions from JAXA and CNES. This is again attributed to almost the entirety of the early work in the field being funded by NASA in the United States.

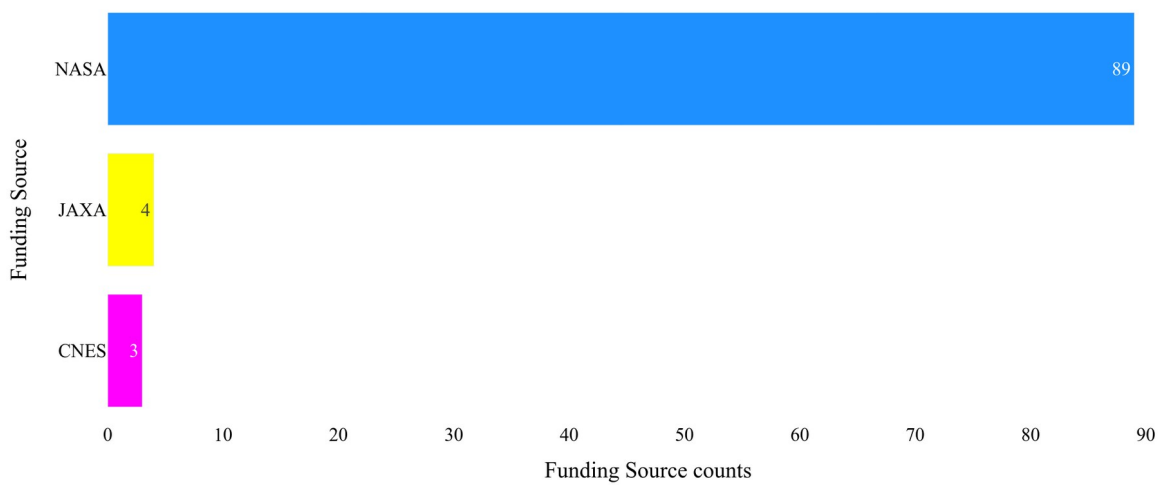

Figure 16. Funding source counts for articles prior to 2013.

Figure 17 shows the top three funding sources (for articles with funding text) after 2013. While NASA continues to be the leading source, its lead over other national sources have decreased. This is attributed to aerocapture not being one of the top funding priorities in the last 10 years, and NASA centers have not had the same level of funding as prior to 2000 to work in this area. The National Natural Science Foundation of China is the next leading source, followed by Charles Stark Draper Lab (USA). As shown in Figure 18, since the mid-2010s, National Natural Science Foundation has been the lead sponsor of aerocapture research in China, a trend which will likely continue in the next decade.

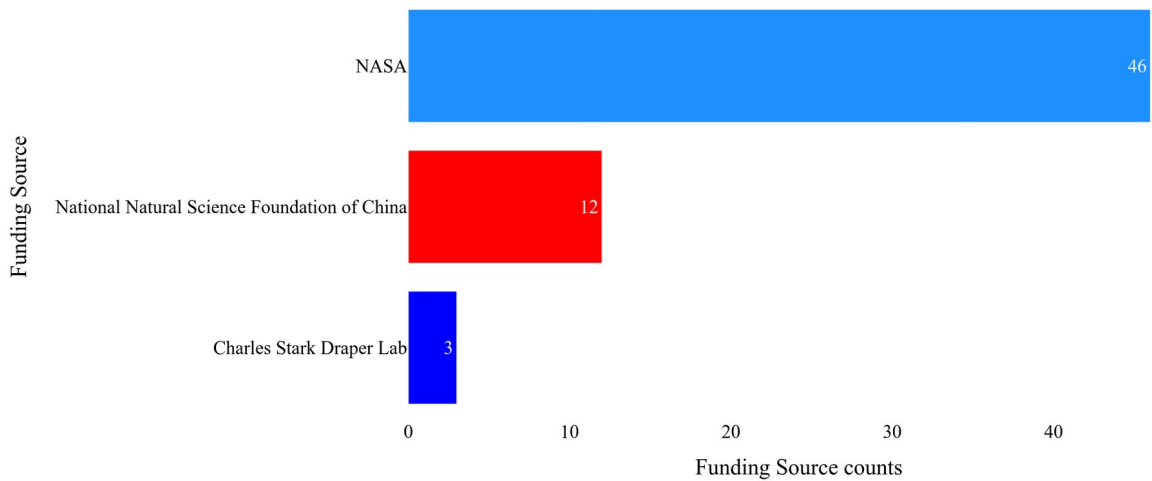

Figure 17 Funding source counts for articles after 2013.

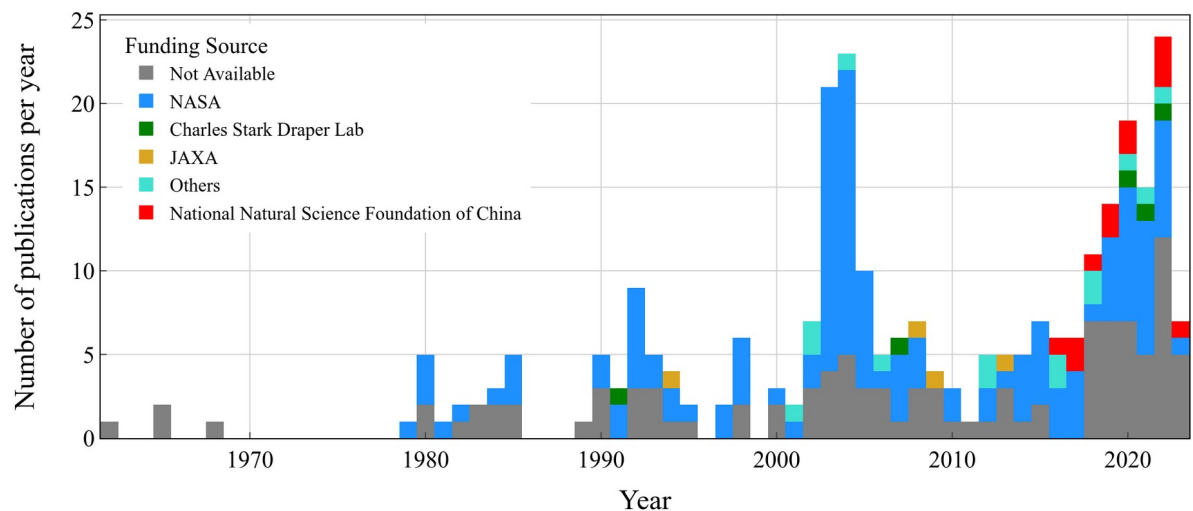

Figure 18. Number of publications per year colored by source of funding.

## IV. FUTURE OUTLOOK

Figure 19 shows a prediction for the cumulative publication count for the next two decades, based on the curve fit for the available data points since the 1980. It is expected the field will continue to grow and will reach 500+ publications by 2043. This prediction is purely 'data-driven', i.e. it assumes the whatever factors such as the rise and fall of funding that had been at work during the last four decades will continue to remain so in the next two decades.

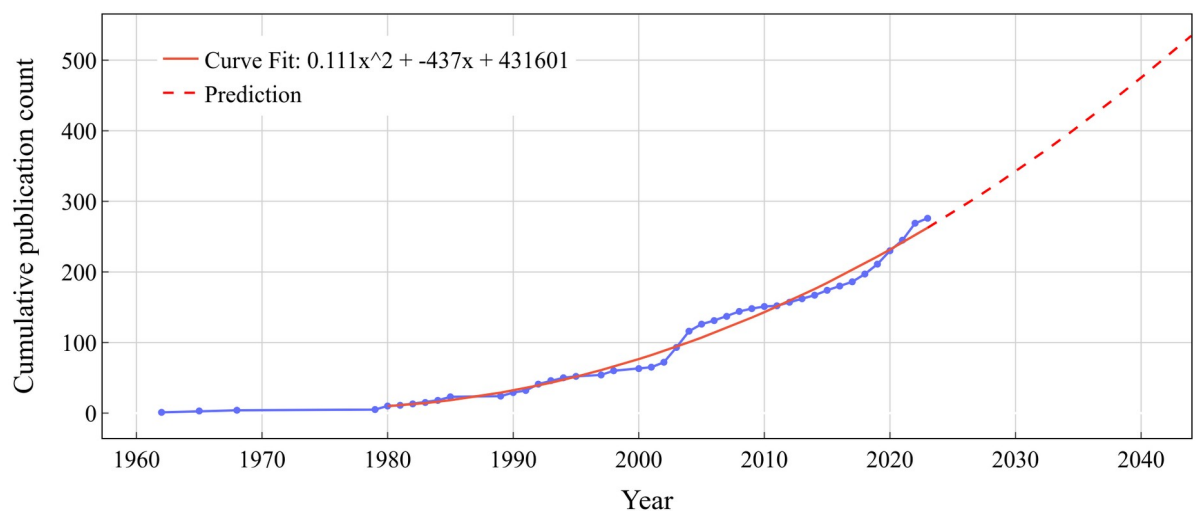

Figure 19. Data-driven prediction for the number of publications in the next 20 years.

However, it is to be noted that while an increasing number of publications certainly contributes to the knowledge in this field, a flight demonstration is needed to take the technology to the next level. This will help it break out of being a 'conceptual' technique and bring it into the realm of a being a 'practical' solution for future missions [32]. The 2023-2032 Planetary Science Decadal Survey notes that 'while aerocapture is a technology ready for mission infusion, because it is not proposed for use in missions, it is considered a "dormant" technology that is perceived as high risk in a mission competitive environment'. The survey further notes that "aerocapture will require special incentives by NASA to be proposed and used in a science mission". In 2022, a review of NASA Space Technology Mission Directorate (STMD) listed an aerocapture flight demonstration as a top priority for future exploration. In presentations given to the Decadal Survey committee by NASA researchers, a plan for a low-cost drag modulation flight demonstration as shown in Figure 20 has been proposed [33]. Such an Earth-based experiment is likely viable within existing programs for technology demonstration, and will greatly reduce the risk for small Mars and Venus missions in the short term [34], and more ambitious Venus missions [35] and outer planet missions in the long term [36, 37, 38].

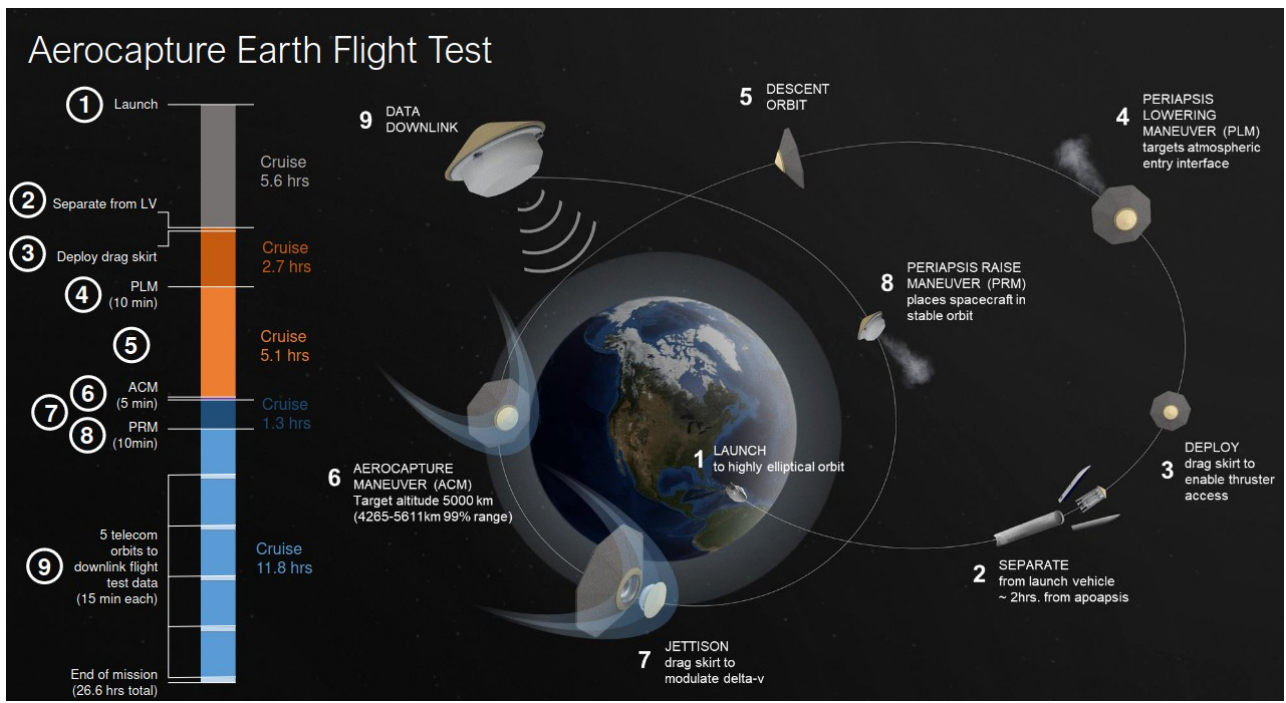

Figure 20. Schematic of the proposed aerocapture Earth flight test demonstration [33].

## V. CONCLUSIONS

The study performed a historical review of the aerocapture concept from its origins since the early 1960s. Bibliometric analysis is performed using data collected from the Web of Science database, and is used to analyze the trends in the publication records from 1980 until 2023. The data reveal a pattern in the rise of publications, followed by a period of stagnation which repeats itself approximately once every decade. Mars is the most studied destination for aerocapture, while Uranus is the least studied. Prior to 2013, NASA centers produced the most publications and are the most cited in the field. However, academic institutions produced the majority of aerocapture publications in the last 10 years.  The United States continues to be the leading country in terms of publications, followed by China.  The Journal of Spacecraft and Rockets is the leading source of aerocapture publications, both in terms of number of publications and citations. NASA continues the leading funding source for aerocapture research, with the National Natural Science foundation funding the majority of aerocapture research in China since the mid-2010s. With only a small amount of government funding, the number of aerocapture publications is expected to continue its slow growth over the next two decades. The 2023-2032 Planetary Science Decadal Survey has identified aerocapture as a technology that is ready for mission infusion, but considered a dormant technology in a cost-competitive scenario. A proposed low-cost Earth flight demonstration of aerocapture will greatly reduce the risk for future science missions.

## DATA AVAILABILITY

The bibliometric dataset used in the study is available at http://dx.doi.org/10.13140/RG.2.2.33580.85121.